\documentclass[twocolumn]{aastex631} % https://journals.aas.org/aastexguide/#style_options

\usepackage{color}
\usepackage{bm}
\usepackage{ amssymb, amsmath }

\newcommand{\g}{\textbf}

\newcommand{\comm}[1]{#1} %commenti
\newcommand{\BB}{\bm{B} }

\newcommand{\bb}{\bm{b} }

\newcommand{\rr}{\bm{r} }

\newcommand{\xx}{\bm{x} }

\renewcommand{\bell}{\boldsymbol{\ell}}

\shorttitle{Evaluation of scale-dependent kurtosis with HelioSwarm}
\shortauthors{Pecora F.}
\graphicspath{{./}}
\begin{document}
\title{Evaluation of scale-dependent kurtosis with HelioSwarm}

\author[0000-0003-4168-590X]{Francesco Pecora}
\affiliation{Department of Physics and Astronomy, University of Delaware, Newark, DE 19716, USA}
\email{fpecora@udel.edu}

\author[0000-0002-5272-5404]{Francesco Pucci}
\affiliation{Consiglio Nazionale delle Ricerche, Istituto per la Scienza e Tecnologia dei Plasmi, (CNR-ISTP), 70126 Bari, Italy}
% francesco.pucci@istp.cnr.it \\

\author[0000-0002-5554-8765]{Francesco Malara}
\affiliation{Dipartimento di Fisica, Università della Calabria, 87036 Rende CS, Italy}
% francesco.malara@fis.unical.it

\author[0000-0001-6038-1923]{Kristopher G. Klein}
\affiliation{Lunar and Planetary Laboratory, University of Arizona, Tucson, AZ 85721, USA}
% kgklein@arizona.edu

\author[0000-0002-5002-6060]{Maria Federica Marcucci}
\affiliation{IAPS - INAF Istituto di Astrofisica e Planetologia Spaziali, 00133 Roma, Italy}
% maria.marcucci@inaf.it

\author[0000-0001-5824-2852]{Alessandro Retin\`o}
\affiliation{Laboratoire de Physique des Plasmas, CNRS, 91128 Palaiseau, France}
% alessandro.retino@lpp.polytechnique.fr

\author[0000-0001-7224-6024]{William Matthaeus}
\affiliation{Department of Physics and Astronomy, University of Delaware, Newark, DE 19716, USA}
% whm@udel.edu



\begin{abstract}
Plasma turbulence involves complex, nonlinear interactions of electromagnetic fields and charged particles across multiple scales. Studying these phenomena in space plasmas, like the solar wind, is facilitated by the intrinsic scale separations and the availability of in situ spacecraft observations. However, the single-point or single-scale configurations of current spacecraft limit our understanding of many properties of the turbulent solar wind. To overcome these limitations, multipoint measurements spanning a range of characteristic scales are essential. This paper prepares for the enhanced measurement capabilities of upcoming multispacecraft missions by demonstrating that higher-order statistics, specifically kurtosis, as a baseline for intermittency can be accurately measured. Using synthetic turbulent fields with adjustable intermittency levels, we achieve scale separations analogous to those in the solar wind and apply these techniques to the planned trajectories of the HelioSwarm mission. This approach promises significant advancements in our understanding of plasma turbulence.
\end{abstract}

%Use showkeys class option if keyword

%\tableofcontents

\section{Introduction}
\label{sec:intro}

Initial observations of atmospheric turbulence by \citet{richardson1922weather} suggested the existence of space-filling eddies of all sizes. This inspired the widely invoked Kolmogorov's theory \citep[][K41]{kolmogorov1941local}. This theoretical development assumed, as one of the fundamental hypotheses of similarity, that the viscosity and the average energy dissipation rate uniquely determine the statistical properties of the turbulent field. However, early results from fluid experiments showed intermittent behavior within the dissipation range evidenced by a nonuniform distribution of energy at small scales \citep{batchelor1949nature}. These observations led to modifications in the theory of turbulence to include the patchy bursts of dissipation in the form of a \textit{locally} averaged energy dissipation rate rather than a \textit{globally} averaged one. This is the so-called refined similarity hypothesis \citep{oboukhov1962some,kolmogorov1962refinement}. After the observation of dissipation-range intermittency, it was also observed in the inertial range with the experimental work of \citet{anselmet1984higherorder} and later confirmed by \citet{gagne1987etude}.

The phenomenology of intermittency, associated with an uneven distribution of energy in space, describes a turbulent medium with a relatively small number of strong, non-space-filling discontinuities, as opposed to the original Richardson description of a large number of uniformly distributed discontinuities \citep{frisch1995turbulence}. This paradigmatic change inspired several theoretical works that aimed at an accurate phenomenological description of higher-order moments that depart from the K41 prescription \citep{mandelbrot1974intermittent,frisch1978simple_betamodel,benzi1984multifractal,meneveau1987simple,she1994universal}. It should be emphasized that these descriptions of intermittency are physically plausible and consistent with observations, but the refined similarity hypothesis itself remains a useful conjecture \citep{wang1996examination,chen1997refined} though still unproven to our knowledge.

Intermittency in the solar wind was first observed by \citet{burlaga1991intermittent,burlaga1991multifractal_b,burlaga1991multifractal_v,burlaga1992multifractal_btn}. Subsequent observations confirmed the earlier findings of intermittent features of the solar wind evidenced by departures of high-order moments from the linear scaling of K41, or, equivalently, as a (more or less pronounced) departure from Gaussianity \citep{marsch1997intermittency, sorriso1999intermittency, padhye2001distribution,voros2002scaling,koga2007intermittent,wan2012investigation,chasapis2020scaling,chhiber2020clustering,roberts2022scale}. Extensions of the hydrodynamic intermittency correction to magnetohydrodynamics (MHD) were developed \citep{carbone1993cascade} together with the possibility of distinguishing between fluid and magnetofluid turbulence \citep{carbone1995experimental}.

 In MHD, and in the solar wind, as in hydrodynamics, the essential physics of intermittency is the observed nonuniformity of the energy density \citep{marsch1996multifractal}. However, again one should note that the theoretical basis for intermittency in MHD remains unproven, as it is in hydrodynamics. Moreover, the precise form of refined similarity in MHD is not well established, see e.g. \citet{merrifield2005scaling}, and remains based on phenomenological arguments (e.g., \citet{ChandranEA15}.)

In MHD, the manifestation of intermittency has been directly associated with the bursts of kinetic energy and dissipation generated at reconnection sites and, more generally, with current sheets and discontinuities \citep{matthaeus1986turbulent,carbone1990coherent,servidio2010statistics,wan2012intermittent,matthaeus2015intermittency}. Novel methods for the detection of intermittent structures were developed based on wavelet analysis  \citep{farge1992wavelet, veltri1999scaling, bruno2001identifying} or the partial variance of increments \citep[PVI,][]{greco2008intermittent,greco2018partial}.

The subsequent applications of these methods originated a vast section of literature that relates intermittent events with kinetic effects such as particle diffusion and heating \citep{osman2012intermittency,osman2012kinetic,tessein2013association,tessein2015effect,chasapis2015thincurrent,mallet2019interplay,gonzalez2024local}, as well as in the macroscopic description of the solar wind as composed of adjacent flux tubes with sharp (discontinuous) boundaries \citep{borovsky2008flux,kittinaradorn2009solar,tessein2016local,pecora2019single_GSPVI,pecora2021identification_HmPVI}.

More recent investigations using the Magnetospheric Multiscale (MMS) mission \citep{burch2016MMS} and Cluster \citep{escoubet2001cluster} have explored similar topics \citep{chhiber2018higher,roberts2022scale}. However, even though the MMS mission provides multipoint measurements, the separations between spacecraft do not allow for multiscale analyses. The methods explored in the present paper extend the methods developed for MMS \citep{chhiber2018higher}. Upcoming missions like HelioSwarm \citep[HS,][]{spence2019helioswarm,klein2023helioswarm} and Plasma Observatory \citep{retino2022particle_PO,marcucci2024PO} aim to rectify these limitations by providing simultaneous measurements over a range of scales.

Simulations, both of MHD and kinetic plasma, have been very useful in elucidating statistical properties of intermittency \citep{biskamp2000scaling, wan2016intermittency, leonardis2016multifractal}. \citet{parashar2015transition} showed that the intermittency level increases with the Reynolds number, highlighting the relevance of system size in such investigations. Results complementary to the present work have been presented recently, quantifying intermittency and employing PVI statistics and magnetic field increments using HelioSwarm-like trajectories \citep{guerreroguio2024exploring}.

The paper is organized as follows. We first provide a brief description of HelioSwarm configuration parameters, Section~\ref{sec:HS}, that will be used in the next Sections. The model used to obtain a turbulent and intermittent magnetic field with scaling properties and ranges of scales similar to the solar wind is described in Section~\ref{sec:model}. The results are presented in Section~\ref{sec:res}. Finally, in the last Section, conclusions are drawn together with a discussion of the implications of these findings for future multispacecraft missions.

\section{HelioSwarm configuration}
\label{sec:HS}

Since at least the 1980 Plasma Turbulence Explorer Study Group Report \citep{montgomery1980report}, the necessity of a multispacecraft observatory for measuring the multiscale characteristics of plasma turbulence has been recognized. Over the past several decades, various concepts for such missions have been proposed, including the CrossScale mission \citep{schwartz2009crossscale}.

The forthcoming HelioSwarm mission \citep{spence2019helioswarm,klein2023helioswarm} features an observatory of nine spacecraft that will provide in situ measurements of near-Earth plasmas, encompassing the magnetosphere and the solar wind. It will provide simultaneous measurements, with inter-spacecraft separations ranging from fluid to ion-kinetic scales. These measurements span a three-dimensional volume to provide adequate coverage of fundamental turbulence statistics \citep[Sec. 4.5.1,][]{klein2023helioswarm}. For direct comparison of orbital parameters with those of the solar wind, we use nominal values at 1 au for the ion skin depth $d_i=100$~km \citep{verscharen2019multi}, and the correlation length $\lambda_c=10^6$~km \citep{matthaeus1999correlation}. These two scales are generally indicative of the extension of the so-called turbulent inertial range.

\begin{figure}[ht]
    \centering
    \includegraphics[width=0.98\columnwidth]{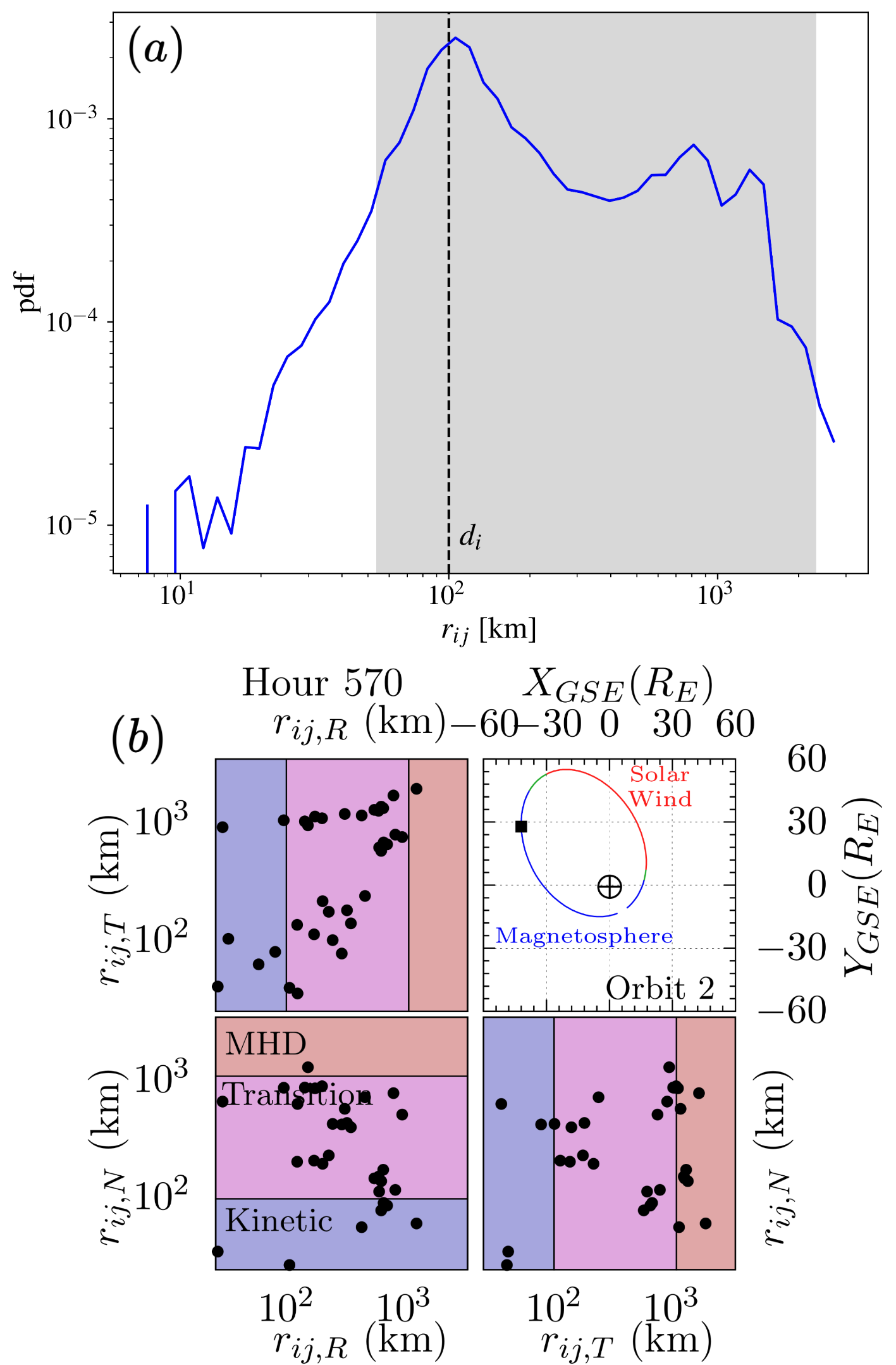}
    \caption{(a) Probability density function (pdf) of HS inter-spacecraft separations throughout the entire duration of the mission. A nominal $d_i$ scale of 100~km is indicated with a vertical dashed line. The configuration used in this work spans the range of scales indicated by the grey-shaded area. (b, upper right) \comm{The black square represents the positions of HS at hour 570 of the planned trajectories (in earth radii, GSE coordinate system)}. The orbit is color-coded by nominal positions of the magnetosphere (blue), pristine solar wind (red), and magnetically connected regions (green). The other three panels illustrate the 36 separations between the 9 spacecraft, projected onto an RTN reference frame. The three stripes in each panel indicate the MHD, transition, and kinetic ranges.}
    \label{fig:pdfrij}
    % 36_HelioSwarm_synth/W_rij_pdf.py
    % https://docs.google.com/presentation/d/1g6htVYhr75mt98CkdkYE20o4d-JgtBVHz27ZTLytAbc/edit#slide=id.p
\end{figure}

The distribution of the range of scales spanned by spacecraft separations, using HelioSwarm Design Reference Mission trajectories, is shown in Fig.~\ref{fig:pdfrij}(a). The interspacecraft separations are measured as $r_{ij} = |\rr_i - \rr_j|$ where $i,j$ is any pair of spacecraft $i,j=1,\dots 9$ and $j>i$. The spacecraft observatory will cover about two orders of magnitude of scale separations across the lower end of the inertial range (marked by $d_i$). For the present work, we have selected hour 570 of the above-mentioned designed trajectories as an example of the observatory having good three-dimensional baseline separations spanning MHD, transition, and ion kinetic scales. At that time, the spacecraft are separated by 54 to 2327~km, as indicated by the shaded area in Fig.~\ref{fig:pdfrij}(a). The position of HelioSwarm and the vector components of the $C_9^2=36$ baselines, projected onto a Radial-Tangential-Normal (RTN) coordinate system are illustrated in Fig.~\ref{fig:pdfrij}(b). As the orbit remains nearly inertially fixed, the apogee undergoes rotation over the course of a year, enabling the observatory to sample the pristine solar wind, the magnetosphere, and areas magnetically connected between these two.
\comm{The HS configuration at hour 570 was selected since it spans over an order of magnitude in scales, making it ideal for examining scale-dependent quantities, such as kurtosis. Moreover, a similar configuration is attained across all three spatial regions (i.e. pristine solar wind, foreshock, and magnetosphere) providing the possibility to perform similar analyses in different environments. For all subsequent analyses, the configuration of the spacecraft remains constant. The virtual spacecraft travel along parallel trajectories, with the fields generated as if they were time series along these paths.}
% The selected hour is representative of good three-dimensional configurations that will be attained in all three of these spatial regions. The HS configuration at hour 570 is kept fixed for all the analyses that follow. The virtual spacecraft fly along parallel lines and the fields are generated along these trajectories (as if they were time series). This configuration covers more than one order of magnitude in scales which is ideal for studying scale-dependent quantities, such as the kurtosis.

\section{Synthetic turbulence model}
\label{sec:model}

The model employed for this work reproduces a synthetic three-dimensional turbulent solenoidal field with nongaussian (intermittent) properties based on the p-model of \citet{meneveau1987simple} as first described in \citet{malara2016fast}. The model does not generate coherent structures but rather statistics of the fluctuations in accordance with that of an intermittent field. Such a field can be directly identified with a magnetic field. The model constructs the magnetic field by superimposing wavelets at various scales from the box size to a desired resolution. The advantage offered by this approach is that at any position, the magnetic field is calculated as the sum of contributions from the wavelets encompassing the point in question. This drastically reduces the number of operations needed to compute the field compared to Fourier methods. The computational time required to evaluate the fields increases linearly with the number $n$ of scales at which the wavelets are defined. The spatial resolution is proportional to $2^{-n}$. This scaling property enables the generation of fields with large spectral bandwidth at low computational costs. Since the wavelets are analytically defined, the field is also analytically defined. Consequently, the field value at any position is computed in real time without the need for a fixed grid. This eliminates the necessity of storing the field in memory, making the algorithm highly suitable for the objectives of this study where we consider a realistic extension of the inertial range. These synthetic turbulent and intermittent magnetic fields have been used for the study of energetic particle diffusion and modeling solar wind fast-speed streams \citep{pucci2016energetic,perri2017estimation,perri2019powerlaw,valentini2019building}. 

%At each iterative step in the building process of the fields, the energy contained within a volume is partitioned so that half of this volume will have a fraction $p$ of the total energy, while the remaining $(1-p)$ is provided to the other half. The larger the value of $p$, the higher the level of intermittency.
\comm{The field is constructed iteratively by partitioning the energy \(\mathcal{E}_{l_m}\) within a volume of scale \(l_m\) following the prescription of the p-model. In this process, half of the subvolume at scale \(l_{m+1}\) receives the fraction \(\mathcal{E}_{l_{m+1}} = 2p \mathcal{E}_{l_m} \left( \frac{l_{m+1}}{l_m} \right)^{2/3}\), while the other half receives \(\mathcal{E}_{l_{m+1}} = 2(1-p) \mathcal{E}_{l_m} \left( \frac{l_{m+1}}{l_m} \right)^{2/3}\). When \(p=1/2\), the two subvolumes at scale \(l_{m+1}\) receive equal energy. For \(p \neq 1/2\), intermittency arises due to the imbalance in energy partitioning.}
This procedure is carried out from the largest (box) scale to a desired minimum resolution. In principle, the domain is periodic, but the periodicity length can be set to be arbitrarily large with respect to the correlation length.

For our purposes, we set the periodicity length (corresponding to the side of the fictitious box in which the fields are defined) to be 80 correlation lengths $(80 \times 10^6~\mbox{km})$. As stated earlier, the fields are not defined on a numerical grid, but instead, they can be evaluated at any desired point through an analytical definition. In order to avoid periodicity effects, we restrain our analyses in a subdomain of side $40 \lambda_c$. We will consider both an isotropic and an anisotropic system.

\begin{figure}[ht]
    \centering
    \includegraphics[width=0.98\columnwidth]{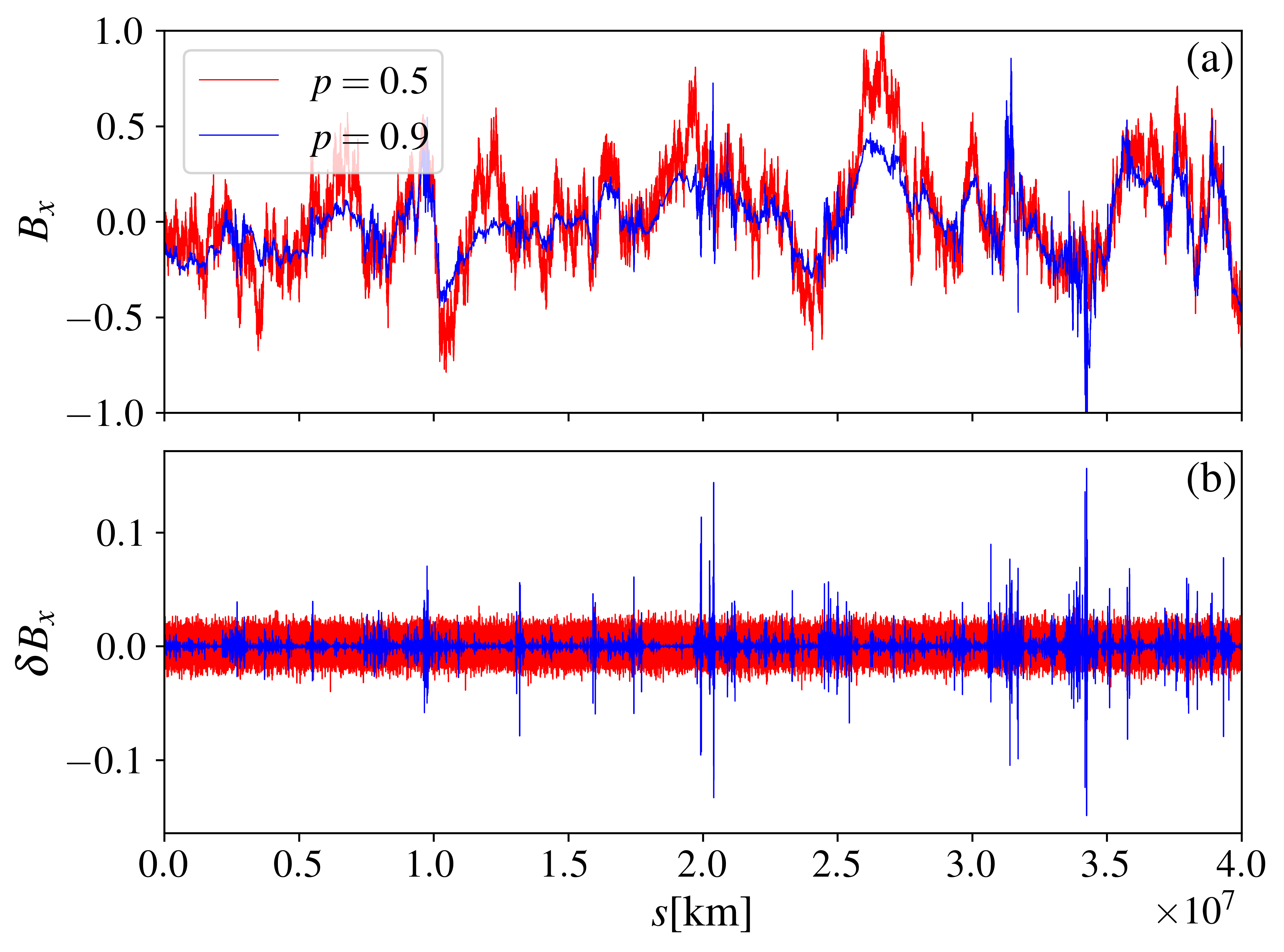}
    \caption{(a) One magnetic field component, $B_x$, for intermittency levels $p=0.5$ (red) and $p=0.9$ (blue) sampled along a line $s$ in any direction (the system is isotropic) with resolution $\delta s \simeq 30$~km. (b) Increments $\delta B_x = B_x(s) - B_x(s+\delta s)$ . The field with no intermittency, $p=0.5$, does not show any ``bursty'' features. The field with higher intermittency, $(p=0.9)$, shows large localized spikes, an indicator of intermittency.}
    \label{fig:dbx}
    % ds = 30.517578125 km
    % G_B_on_traj_v*.py
\end{figure}

For the former, the smallest scale is set to correspond to the ion inertial length in the solar wind at 1~au \citep[$d_i=100$ km,][]{verscharen2019multi}. The spacecraft ``time series'' have a sampling resolution of 30~km (or, equivalently a 75~ms with a nominal solar wind speed of 400~km/s).  We generated three sets of magnetic fields with increasing levels of intermittency. The field with $p=0.5$ has Gaussian increments (no intermittency). The other two fields with $p=0.7$ and $p=0.9$ show increasing levels of intermittent behavior as seen in Fig.~\ref{fig:dbx} for the two extreme cases of $p=0.5$ and $p=0.9$. In the Figure, we show the magnetic field component $B_x$ obtained on a line sampling $40\lambda_c$ along a generic direction $s$.
For the anisotropic case, we focus on the intermittency level $p=0.7$ and, for reasons detailed below, the smallest scale is set to 0.015~km and the resolution of the time series is 60~km (or 150~ms).

\begin{figure}[ht]
    \centering
    \includegraphics[width=0.98\columnwidth]{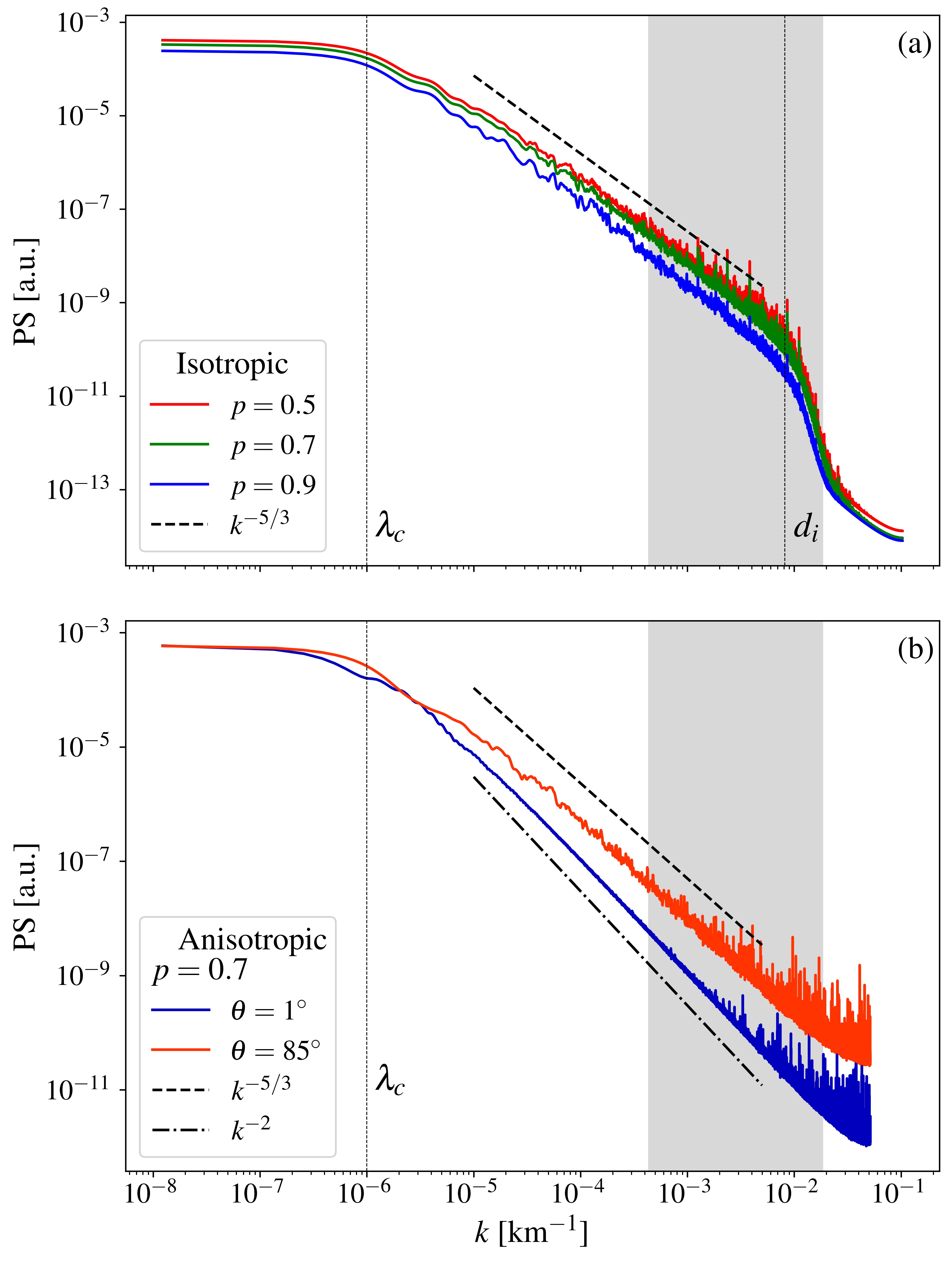}
    
    \caption{Magnetic field power spectra computed over single spacecraft trajectories and then averaged for (a) isotropic and (b) anisotropic cases. $k^{-5/3}$ and $k^{-2}$ dashed and dot-dashed lines are added as reference. The gray-shaded area indicates the range of scales covered by the HS separations.
    In (a) it is evident that higher intermittency $(p=0.9)$ generates slightly steeper spectra, though the extension of the inertial range remains intact. Correlation and ``dissipation'' scales are indicated with vertical dotted lines.
    In (b) the spectra are computed when the HS formation flies at angles $\theta=1^\circ$ and $\theta=85^\circ$ with respect to the mean field direction. The apparent flattening is only due to a downsampling of the measurements. The spectra would have a decay similar to that occurring near $d_i$ in (a) at $k\sim 67$. The slopes have also been estimated using a fit in the region indicated by the reference lines obtaining -1.91 and -1.64 for $\theta=1^\circ$ and $\theta=85^\circ$ respectively.}
    \label{fig:psd}
    % 36_HelioSwarm_synth/G_Sp_v02.py

    % RICORDA DI DIRE CHE LA SCALA DISSIPATIVA È GIÙ GIÙ DA QUALCHE PARTE
    % theta=5,  slope = -1.68011379
    % theta=85, slope = -1.64237961
    % 36_HelioSwarm_synth/G_Sp_aniso_v02.py
\end{figure}

The three sets of fields, for the isotropic case, have been generated imposing a nominal spectral slope of $-5/3$ within the inertial range as is frequently observed in the solar wind \citep{kiyani2015dissipation}. Below the chosen smallest scale ($d_i$), the spectrum decays exponentially. The field with $p=0.9$ shows a slightly steeper slope as higher levels of intermittency are expected to modify the slope of the power spectrum \citep{oboukhov1962some,kolmogorov1962refinement}. Figure~\ref{fig:psd}(a) shows the power spectra for the three fields (with the different intermittency levels); indicated are the scales corresponding to $\lambda_c$ and $d_i$, and the range of scales covered by HelioSwarm for the selected Design Reference Mission hour as a shaded area.

To investigate possible effects due to anisotropy, a case relevant to the solar wind, we obtain a realization of fields with different spectral slopes in the directions parallel and perpendicular to the mean magnetic field, with a fixed $p=0.7$. The anisotropy is obtained by elongating the support of the wavelets in the direction parallel to the mean magnetic field, as explained in \citet{malara2016fast}. As a result of that procedure, the parallel spectrum decays exponentially at scales larger than that in the perpendicular direction. To ensure that the spectrum in the parallel direction does not fall off exponentially in the range of scales covered by HelioSwarm, we shifted the onset of the exponential decay to these smaller scales.

In this realization, we fly the HS observatory at 8 different angles $\theta$ ranging from $1^\circ$ to $85^\circ$ with respect to a mean field $B_0 \hat{z}$. In Fig.\ref{fig:psd}(b) we show the magnetic field power spectra and, as in (a), the correlation length is indicated with a vertical line, and the scales spanned by the observatory are represented by the grey-shaded area. In this case, the spectra do not show the exponential falloff as they extend to k's larger than those shown in Fig.~\ref{fig:psd}(a). The smallest scale in this case is 0.015~km which correspond to $k \sim 67$. The apparent flattening of the spectra at large k's is due to downsampling of the fields to obtain reasonable sizes of the data arrays.

% \begin{figure}[ht]
%     \centering
%     \includegraphics[width=0.5\columnwidth]{figs/PS_ave_saniso.png}
%     \caption{Power spectra of the magnetic field for the two cases in which the HS formation flies at an angle $\theta=1^\circ$ and $\theta=85^\circ$ with respect the mean field direction. The scales spanned by the spacecraft are indicated by the gray-shaded region. The vertical dotted line indicates the correlation scale. Slope for $\theta=1$, -1.91, for $\theta=85$, -1.64. Reference lines for $k^{-5/3}$ (dashed) and $k^{-2}$ (dash-dotted) are present.}
%     \label{fig:sp_ani}
%     % RICORDA DI DIRE CHE LA SCALA DISSIPATIVA È GIÙ GIÙ DA QUALCHE PARTE
%     % theta=5,  slope = -1.68011379
%     % theta=85, slope = -1.64237961
%     % 36_HelioSwarm_synth/G_Sp_aniso_v02.py
% \end{figure}

\section{Results}
\label{sec:res}

Formally, intermittency is the deviation of the distribution of some quantity from a Gaussian distribution. This deviation is expressed by extreme values that are more probable than those observed in a normally distributed population. One way to quantify this is to compare the fourth-order to the second-order moment squared of the scalar increments $\delta f$, defined as the difference $f(\xx) - f(\xx+\bell)$ between two points in space separated by a (vector) lag $\bell$. We use the longitudinal increments of the magnetic field $\delta b_{\ell} = \delta \bb \cdot \hat{\bell} =  ( \BB_i(\xx_i) - \BB_j(\xx_j) ) \cdot \hat{\bell} $ where $\BB_i(\xx_i)$ is the magnetic field measured at position $\xx_i$, $\BB_j$ is at position $\xx_j = \xx_i + \bell$, and $\hat{\bell}$ is the unit vector along the $\bell$ direction. To achieve the comparison, we operationally define the scale-dependent kurtosis as

\begin{equation}
    \kappa(\bell) = \frac{ \langle \delta b_\ell^4 \rangle  }{ \langle \delta b_\ell^2 \rangle^2 }
    \label{eq:k}
\end{equation}
where $\langle.\rangle$ indicates an ensemble average \citep{frisch1995turbulence}. When the increments of a scalar field follow a Gaussian distribution, $\kappa(\bell)=\kappa=3$. The presence of extreme fluctuations produces values of $\kappa$ larger than 3. We measure the scale-dependent kurtosis $\kappa$ by adapting the definition in Eq.~\ref{eq:k} to four different scenarios. 

\begin{figure}[ht]
    \centering
    \includegraphics[width=0.98\columnwidth]{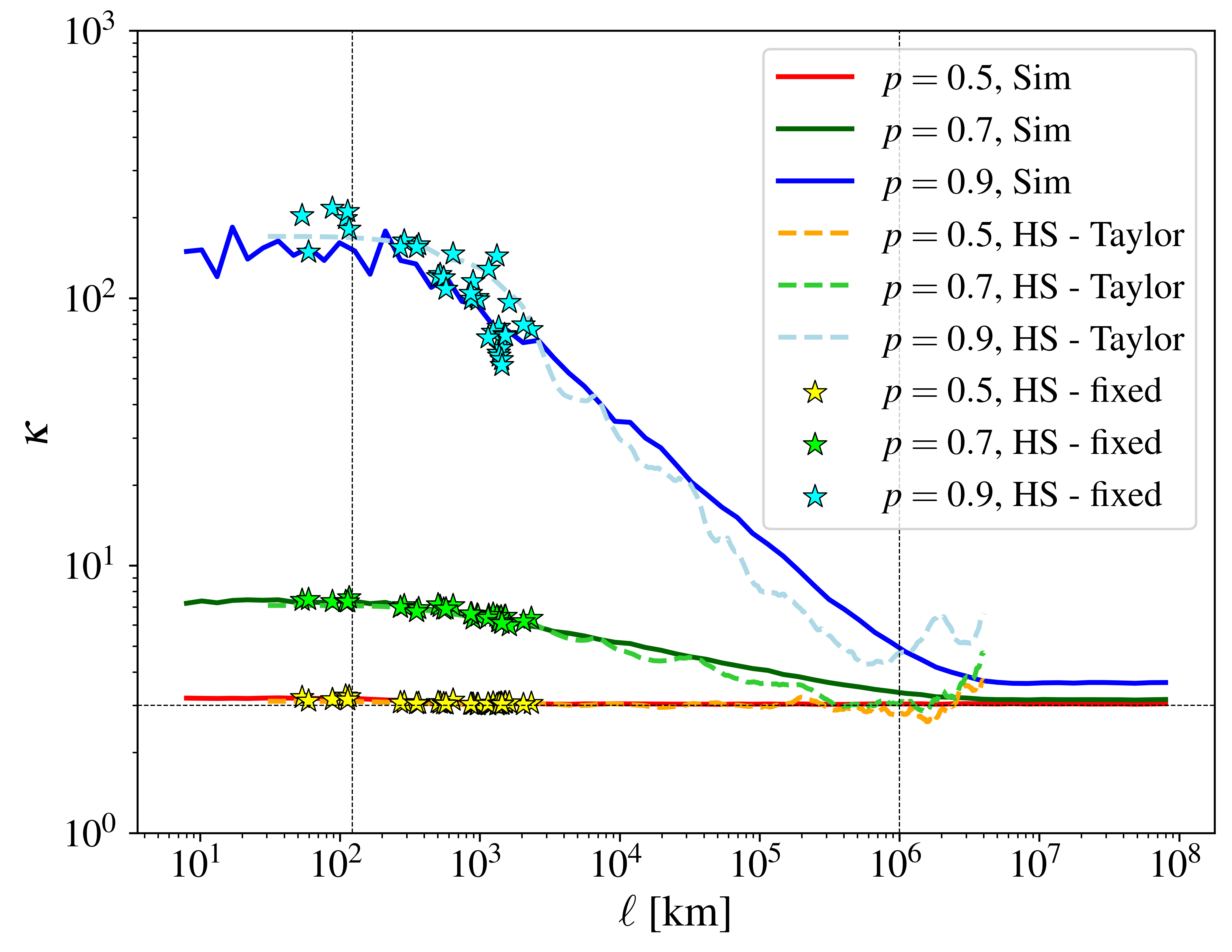}
    \caption{Kurtosis of the longitudinal magnetic field increments. Solid lines are obtained from millions of random paired points scattered throughout the whole volume -- see text, case (i). Dashed lines result from a Taylor-like approach on single-spacecraft trajectories, then averaged, case (ii). Symbols are obtained from fixed inter-spacecraft separations, case (iii).}
    \label{fig:kurt}
    % G_kurt_v*.py
\end{figure}

(i) We first need the accurate kurtosis obtainable from the entire sample of the synthetic fields. These are the data to which the spacecraft sample results are going to be compared. Since the fields are not defined on a grid, we choose several million pairs of points (over which the ensemble average of Eq.~\ref{eq:k} are calculated) at random positions within the domain. Separations are randomly oriented with magnitudes equally spaced, ranging from the domain size ($8\times10^7$~km) to 8~km. This approach provides the solid lines in Fig.~\ref{fig:kurt} that are representative of the kurtosis of the synthetic fields over the whole domain.

For the next three approaches, we use the HelioSwarm trajectories.

\comm{(ii) To evaluate the scale-dependent kurtosis, we use spatial increments along the trajectory of each spacecraft. These increments are represented by $\delta \bb_{\bell} = (\BB_i(\xx_i) - \BB_i(\xx_i + \bell_i))$, where $\bell_i$ denotes the spatial increments along the trajectory of spacecraft $i$. This method is similar to measuring temporal increments in a spacecraft's time series and then applying the Taylor hypothesis exactly, which assumes that the structures observed in the flow are ``frozen’’ and advected past the spacecraft without changing shape \citep{taylor1938spectrum_frozenin}. The ensemble average is taken along each trajectory and then we averaged the resulting $\kappa$ over the 9 trajectories for additional statistical weight. Good agreement of the HelioSwarm-Taylor-like scale-dependent kurtosis with the validated one of approach (i) is attained within the range of scale up to $1/10$ of the correlation length.}

(iii) This approach fully exploits  HelioSwarm's multi-point potential. The scale-dependent nature of kurtosis is analyzed without the need for Taylor's hypothesis, thus avoiding 
problems of space-time correlation, especially at smaller scales. Instead, the increments are computed at the fixed separations between the spacecraft, $\delta \bb(\bell_{ij}) = ( \BB_i(\xx_i) - \BB_j(\xx_j) )$ where $\BB_i(\xx_i)$ is the magnetic field at the position $\xx_i$ of the $i$-$th$ spacecraft, $\bell_{ij}=\xx_j-\xx_i$, and the ensemble average is taken over the whole trajectory. The observatory of 9 spacecraft allows for 36 values of kurtosis to be computed with different lags; these are 
shown by the symbols in Fig.~\ref{fig:kurt}.

\comm{(iv) Finally, we analyzed the effects of increments measured between different spacecraft, as discussed in previous studies \citep{horbury2000cluster,osman2011anisotropic}. Similar to strategy (ii), we assume the validity of the Taylor hypothesis. Using this "modified" Taylor hypothesis, the magnetic field increments measured across different spacecraft time series are defined as $\delta \bb(\bell_{ij}) = (\BB_i(\xx_i) - \BB_j(\xx_j + \bell_j))$, where $\bell_j$ represents the spatial increments along the trajectory of spacecraft $j$. Contrary to case (iii), this approach results in 36 lines instead of 36 points. The outcomes from this method are nearly identical to those obtained in the previous cases and are therefore not displayed below.}

All methods show high levels of agreement, regardless of the intermittency levels. As expected, the field with Gaussian increments $(p=0.5)$ shows a constant value of $\kappa=3$ across all scales. Of particular relevance is the accuracy of the scale-dependent kurtosis obtained by the spacecraft using their fixed separations without the employment of the Taylor hypothesis. The results shown in Fig.~\ref{fig:kurt} are obtained using the isotropic fields. In the following, we investigate the effects of anisotropy on the evaluation of scale-dependent kurtosis.

For the analysis of the scale-dependent kurtosis in an anisotropic medium, we use approaches (i) and (iii) described above. Approach (i) is used to show agreement of the results with the global statistical properties of the fields and (iii) exploits the unprecedented capabilities of novel multiscale missions. In the latter case, two different angles need to be defined as will be relevant below. $\theta$ is the angle of the (parallel) trajectories with respect to $\hat{z}$. $\varphi$ is the angle between the vector separations $\rr_{ij}$ with $\hat{z}$.

% \begin{figure}[ht]
%     \centering
%     \includegraphics[width=0.2\columnwidth]{figs/fig_theta_phi.png}
%     \caption{Schematic representation of the two angles $\theta$ and $\varphi$. The former is the angle that any trajectory (dotted line) forms with the anisotropic direction $\hat{z}$, and the latter is the angle formed by the interspacecraft vector separation.}
%     \label{fig:thetaphi}
%     % https://docs.google.com/presentation/d/1Myok13PBu6BdoP4XOPaSZg7GqOtK0jzJL4Epbw_jdZs/edit#slide=id.p
% \end{figure}

The ``reference'' scale-dependent kurtosis obtained using method (i) for angles ranging from $\theta=1^\circ$ to $\theta=85^\circ$ is illustrated in Fig.~\ref{fig:kurt}(a). Higher levels of intermittency are observed for directions parallel to the mean field. This can be understood in the context of the present model as follows. The anisotropy arises from ``splitting'' the energy content of a subvolume into subdomains elongated in a preferred direction, as described by \citep{malara2016fast}. This elongation of subdomains implies that the same value of kurtosis seen in the perpendicular direction is achieved in the parallel direction at much larger scales. Furthermore, it is possibly reasonable to expect higher kurtosis in this direction, considering the heuristic interpretation of kurtosis as a measure of the inverse of the filling factor of the intermittent features \citep{frisch1995turbulence}.

\begin{figure}[ht]
    \centering
    \includegraphics[width=0.9\columnwidth]{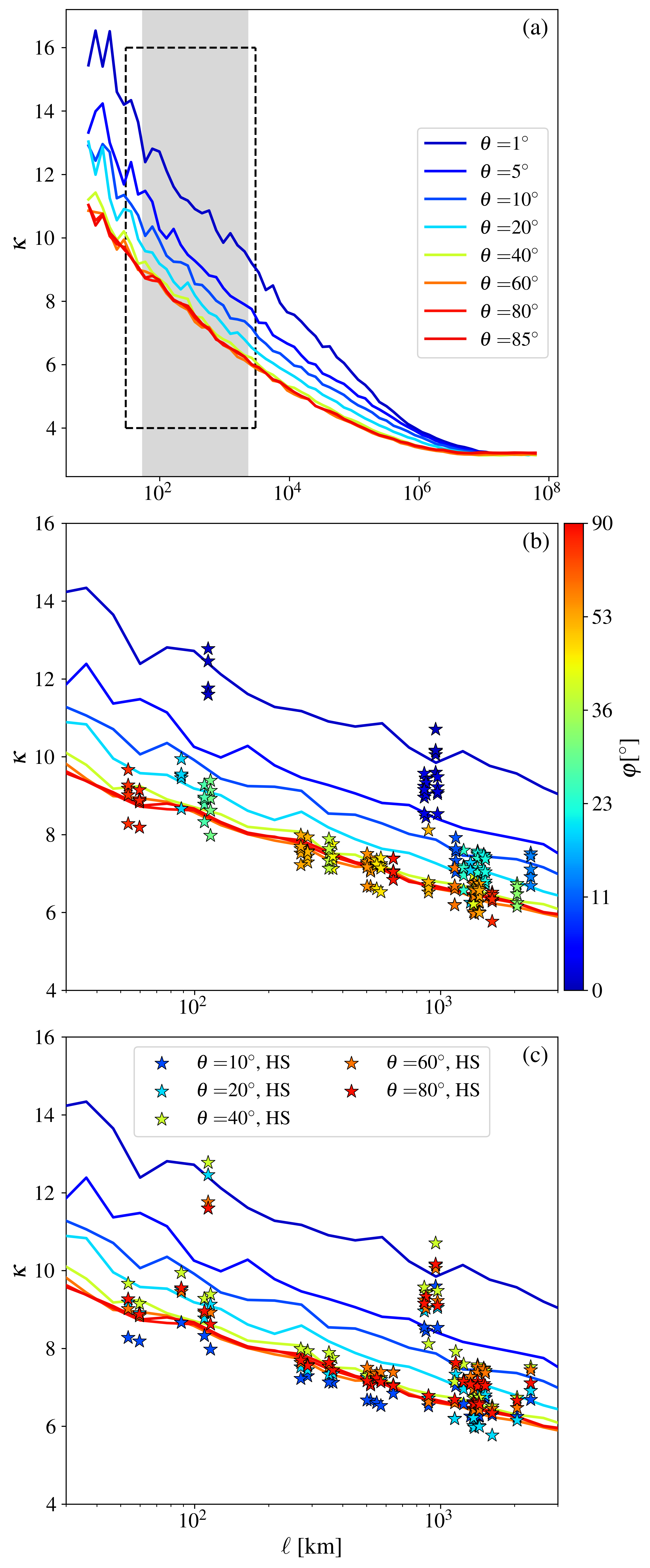}
    \caption{(a) Kurtosis obtained from the anisotropic fields in different directions $\theta$ with respect to the mean field using method (i). Indicated is the range of scales spanned by the spacecraft as a gray-shaded area. The dashed box indicates the region focus of (b) and (c). In (b) the HS $\kappa$ values (symbols, method (iii)) are colored by the angle $\varphi$ the vector separation $\rr_{ij}$ makes with the $\hat{z}$ direction. In (c), the stars are colored by the angle $\theta$ at which the observatory flies with respect to $\hat{z}$. Notice the good agreement of method (iii) with the expected values when the angle $\varphi$ is considered.}
    \label{fig:kurt_ani_sim}
    % 36_HelioSwarm_synth/G_kurt_saniso_v04.py
\end{figure}

We then explored the accuracy of determining the scale-dependent kurtosis using the HS pairs, as outlined in strategy (iii). In this case, along with considering the angle $\theta$ formed by the trajectory with the mean field direction $\hat{z}$, we also need to account for the angle $\varphi$ formed by the vector separation $\rr_{ij}$ with respect to $\hat{z}$. As depicted in Fig.~\ref{fig:kurt}(b), where the symbols are color-coded by the angle $\varphi$ that each HS pair forms with $\hat{z}$, it becomes apparent that the spacecraft ability to capture different kurtosis values hinges not only on their relative distances but also on their orientation. Conversely, neglecting the pair's orientation with respect to the mean field, as in (c) where the symbols are color-coded with the angle $\theta$ formed by the trajectory with $\hat{z}$, leads to less precise kurtosis measurements. Although the general scale-dependent trend remains discernible, certain values deviate due to the pairs sampling diverse orientations relative to the mean field.

To maintain clarity, fewer $\theta$ trajectories are displayed in Fig.~\ref{fig:kurt_ani_sim}(b) and (c), as their overall behavior remains consistent. Only the trajectory at $\theta=1^\circ$ tends to slightly overestimate kurtosis values. This can be rationalized by considering that an almost perfectly parallel trajectory, when no dynamical activity is present, would register smaller variations along its path. Consequently, the spacecraft essentially measure similar kurtosis values along their trajectory as those observed at their initial position, making it unrepresentative of the entire domain.

\section{Conclusions}

In the new era of multipoint multiscale space missions, it is imperative to enhance our management of the increasing volume and complexity of data. In this study, we presented an analysis of the remarkable potential of the forthcoming HelioSwarm mission in measuring scale-dependent kurtosis. We demonstrated that the HelioSwarm observatory not only provides adequate coverage for accurate convergence of this higher-order statistic but also allows for a comprehensive investigation of its nature in an anisotropic medium such as the solar wind.

We employed a model for generating synthetic turbulence fields that does not rely on a fixed grid, enabling the investigation of a range of spatial scales similar to those of the solar wind. The analyses were performed in a three-dimensional domain of size $80 \times \lambda_c = 8 \times 10^7$~km per side, with characteristic ``dissipation'' scales (scale at which the power spectra have an exponential decay) $d_i = 100$ km and 0.015~km for the isotropic and anisotropic systems, respectively.

We selected a time in the current design reference mission trajectories of HelioSwarm when the spacecraft had separations ranging from 54 to 2327~km. This specific configuration, which spans more than one order of magnitude in scales, is optimal for analyzing scale-dependent quantities such as kurtosis.

Initially, we assessed the accuracy of kurtosis estimates by HelioSwarm pairs in an isotropic medium. We compared the scale-dependent kurtosis obtained using four different strategies. (i) We used millions of pairs with random orientations and increasing separations to measure scale-dependent kurtosis homogeneously throughout the entire domain from 8 to $8 \times 10^8$~km. This measure serves as a reference for the other methods. (ii) We employed magnetic field measurements along single-spacecraft trajectories and applied the Taylor hypothesis to compute the kurtosis. For the isotropic case, this measure (averaged over all nine trajectories) provided an accurate prediction for scales up to 1/10 of the correlation length, with deviations likely due to the lower statistical weight of the largest lags. (iii) We used fixed separations between spacecraft pairs to evaluate kurtosis, and the HelioSwarm observatory allowed for up to 36 independent estimates, reported to be accurate regardless of the intermittency level. (iv) We applied a modified Taylor hypothesis, computing increments across different spacecraft pairs. This method also agreed with the others (not shown), demonstrating the robustness of the analyses applicable to the anisotropic case.

Next, we introduced spectral anisotropy of the fields, enforcing a steeper power spectrum for the magnetic field in a specified direction (e.g., $\hat{z}$). We obtained a magnetic field with a spectrum following a $k^{-5/3}$ power law in the perpendicular plane (relative to $\hat{z}$) and $k^{-2}$ in the parallel direction. We compared results from methods (i) and (iii). From (i), we observed that scale-dependent kurtosis had higher values when sampled in the parallel direction, reflecting the model-generated anisotropy. Using method (iii), there are at least two possible ways to analyze the results in the anisotropic medium. If results are cast in a fashion similar to the isotropic case, one perceives that the overall scale-dependent trend is somewhat maintained even though non-negligible variations are present (Fig.~\ref{fig:kurt_ani_sim}(c)). The nature of these variations is unveiled when the values are associated with the angle formed by the spacecraft pair with the preferred direction (Fig.~\ref{fig:kurt_ani_sim}(b)). In this case, an order is recovered and it shows that the ``correct'' value of kurtosis is obtained when the orientation of the pair is taken into consideration more so than with respect to the orientation of the trajectory itself. 

The results obtained in this work using a synthetic model for a turbulent magnetic field with scale separations nominally identical to those of the solar wind further support the effectiveness of multipoint multiscale missions for the observation of quantities as not previously possible in space plasmas. The general accordance and robustness of the results preannounces reliable scale-dependent measures in the solar wind. 
This work contributes to the growing body of research dedicated to fully exploiting future multispacecraft data \citep{maruca2021magnetore,pecora2023helioswarm,broeren2024multisc} in preparation for missions such as HelioSwarm \citep{spence2019helioswarm,klein2023helioswarm} and Plasma Observatory \citep{retino2022particle_PO,marcucci2024PO}.

\begin{acknowledgments}
This work is supported by NASA MMS Mission under grant number 80NSSC19K0565 at the University of Delaware, and by the HelioSwarm Mission.  
FP acknowledges the project ``Data-based predictions of solar energetic particle arrival to the Earth: ensuring space data and technology integrity from hazardous solar activity events''
(CUP H53D23011020001) Finanziato dall’Unione europea – Next Generation EU PIANO NAZIONALE DI RIPRESA E RESILIENZA (PNRR) Missione 4 ``Istruzione e Ricerca'' - Componente C2 Investimento 1.1, ``Fondo per il Programma Nazionale di Ricerca e Progetti di Rilevante Interesse Nazionale (PRIN)'' Settore PE09.
We acknowledge EuroHPC JU for awarding project access to MeluXina CPU at LuxProvide (LU).
\end{acknowledgments}
% \clearpage
% a
% \newpage

%\appendix

%\section{appendice 1}

%\end{multicols}

%\nocite{*}

%\bibliography{biblio_ciccio,refs_WHM}% Produces the bibliography via BibTeX.

\end{document}